\documentclass[a4paper,fleqn,usenatbib]{mnras}

\usepackage{mathptmx}

\usepackage[T1]{fontenc}
\usepackage{ae,aecompl}

\newcommand*{\thead}[1]{\multicolumn{1}{c}{#1}}
\usepackage{graphicx}   
\usepackage{amsmath}    
\usepackage{amssymb}    

\usepackage{comment}
\includecomment{versiona}

\usepackage[usenames]{color}

\usepackage[normalem]{ulem}

\newcommand{\be}{\begin{equation}}
\newcommand{\ee}{\end{equation}}
\newcommand{\bi}{\begin{itemize}}
\newcommand{\ei}{\end{itemize}}

\title[Weighted density fields as gravity probes]{Weighted density fields as improved probes of modified gravity models}

\author[Llinares and McCullagh]{
Claudio Llinares,$^{1}$\thanks{E-mail: claudio.llinares@durham.ac.uk}
and Nuala McCullagh$^{1}$
\\
$^{1}$Institute for Computational Cosmology, Department of Physics, Durham University, Durham DH1 3LE, U.K.
}

\date{Accepted XXX. Received YYY; in original form ZZZ}

\pubyear{2017}

\begin{document}
\label{firstpage}
\pagerange{\pageref{firstpage}--\pageref{lastpage}}
\maketitle

\begin{abstract}
When it comes to searches for extensions to general relativity, large efforts are being dedicated to accurate predictions for the power spectrum of density perturbations.  While this observable is known to be sensitive to the gravitational theory, its efficiency as a diagnostic for gravity is significantly reduced when Solar System constraints are strictly adhered to.  We show that this problem can be overcome by studying weigthed density fields.  We propose a transformation of the density field for which the impact of modified gravity on the power spectrum can be increased by more than a factor of three.  The signal is not only amplified, but the modified gravity features are shifted to larger scales which are less affected by baryonic physics.  Furthermore, the overall signal-to-noise increases, which in principle makes identifying signatures of modified gravity with future galaxy surveys more feasible.  While our analysis is focused on modified gravity, the technique can be applied to other problems in cosmology, such as the detection of neutrinos, the effects of baryons or baryon acoustic oscillations. 
\end{abstract}

\begin{keywords}
gravitation -- cosmology:theory -- cosmology:dark energy -- cosmology:dark matter -- cosmology:large-scale structure of Universe
\end{keywords}



\section{Introduction: weighted density distributions}

The discovery in 1998 of the acceleration of the redshift-luminosity relation of galaxies led to a revolution in our view of the Universe \citep{1998AJ....116.1009R, 1999ApJ...517..565P}.  Within the standard paradigm, these observations are interpreted as an acceleration of the expansion rate of the Universe, which is ascribed to an unknown form of energy called ``dark energy''.  The same observations led some theorists to explore alternative theories of gravity, beyond Einstein's theory of general relativity (GR) \citep{2012PhR...513....1C}.  While predictions from modified gravity (MG) models can differ widely from GR, there are strict constraints that come from the solar system and the cosmic microwave background (CMB).  Solar System observations can be reproduced within a MG context by introducing screening mechanisms, which force the modifications to be suppressed at very small scales and in high-density regimes. Several screening mechanisms exist, including:  Vainshtein, symmetron, chameleon, disformal and D-Bionic \citep[][]{vain, 2010PhRvL.104w1301H, cham, 2012PhRvL.109x1102K, Burrage:2014uwa}.  While they can reconcile alternative gravity theories with data, they significantly reduce the impact of MG on the large scale structure of the Universe, making it difficult to detect deviations from GR using standard cosmological observables such as the power spectrum of density perturbations \citep{2013MNRAS.428..743L, 2015MNRAS.449.2837G}.  This letter proposes a method to overcome this problem.

One reason 2-point statistics are insensitive to MG when the Solar System constraints are taken into account is that the signal is dominated by high density peaks.  In these regions, the screening is active and thus MG effects are minimized.  A possible workaround consists of pruning the data from these high density peaks, a procedure known as clipping \citep{2011PhRvL.107A1301S, 2015PhRvL.114y1101L}.  A similar strategy involves reducing the importance of the high density peaks by making a log transformation of the density before computing the power.  The use of the log transform is not new in cosmology.  As late time density fields are highly non-Gaussian, the 2-point statistics do not contain all the information.  Information can be recovered by making the fields closer to Gaussian, which can be done through the log transform \citep[e.g.][]{2009ApJ...698L..90N, 2011ApJ...731..116N, 2011ApJ...742...91N, 2013ApJ...763L..14M, 2016MNRAS.457.3652M}.  The Gaussianisation of the fields increases the signal-to-noise of the observable and thus its ability to disentangle between different models.

While the weighting associated with the log transform can improve the detectability of MG, there is no reason to assume it is optimal.  We propose a generalization of the log transform (the restricted log trasform), which picks out a specific density range to up-weight, and down-weights all other densities in a continuous way.  Thus, the three density transformations we consider are:
\begin{align}
\label{change_0}
f_0(\rho) & = \rho/\rho_0 \\
\label{change_1}
f_1(\rho) & = \log_{10}(\rho/\rho_0) \\
\label{change_2}
f_2(\rho; A, B, C) & = \left(\log_{10}(\rho/\rho_0)+A\right) \\
\nonumber
&~~~~~~~~~~~\exp\left[-(\log_{10}(\rho/\rho_0)-B)^2/(2C)^2 \right], 
\end{align}
where $\rho$ and $\rho_0$ are the perturbed and mean matter densities.  These three functions correspond to the usual density and the log transform ($f_0$ and $f_1$) and to the restricted log transformation proposed here ($f_2$).  The three free parameters in the restricted log transform ($A$, $B$ and $C$) select the domain of densities which are explored in log space and will be chosen to maximize both the difference between gravity models and the signal-to-noise of the measurements.  We will test the ability of these transformations to disentangle between different models by comparing predictions obtained from N-body simulations that were run with GR and two different MG models.  For simplicity, we will show results at redshift $z=0$.

\section{The gravitational models and simulations}

We consider two alternative gravity models, which were constructed to give an alternate explanation of the accelerated expansion of the Universe and are representative of two different screening mechanisms.  The aim is to show that the density transformations proposed above are capable of distinguishing between GR and MG as well as between these two models.  These models -- symmetron and a specific realization of $f(R)$ -- correspond to the family of scalar tensor theories and have been widely studied in a cosmological context \citep[e.g.][]{2011PhRvD..84j3521H, 2014AnP...526..259L}.  Here we summarize a few specifics, as well as describe the set of simulations that was used for the analysis.  See \citet{2014A&A...562A..78L} for more details on the models and their implementation into the N-body code used for the cosmological simulations.

The dynamics of the symmetron field $\phi$ \citep{2010PhRvL.104w1301H} is defined by the following action
\be
S = \int \sqrt{-g} \left[ R - \frac{1}{2}\nabla^a\phi \nabla_a \phi - V(\phi)\right] d^4x + S_M(\tilde{g}_{ab}, \psi), 
\ee
where $S_M$ is the matter action and $R$ is the Ricci scalar associated to the metric $g_{ab}$.  The geometry is given by the Einstein's frame metric $g_{ab}$ and the matter fields $\psi$ are coupled to the Jordan frame metric $\tilde{g}_{ab}$, which is conformally related to $g_{ab}$.  The potential $V$ and conformal factor that relates the two metrics are chosen in such a way that the scalar field behaves as a usual scalar field oscillating in the following effective potential:
\be
V_{\mathrm{eff}}(\phi) = \frac{1}{2}\left(\frac{\rho}{M^2} - \mu^2\right)\phi^2 + \frac{1}{4}\lambda\phi^4 + V_0,
\ee
where $\mu$, $M$, $\lambda$ and $V_0$ are free parameters.  In regions where the density is low, the scalar field oscillates away from zero giving rise to an enhancement of gravity called fifth force.  When the density increases above a given value, the shape of the potential changes, forcing the scalar field to oscillate around zero and thus to be screened.  For numerical convenience, we substitute the original free parameters of the model $(M, \mu, \lambda)$ with $(\lambda_0, a_{SSB}, \beta)=(1/(\sqrt{2}\mu), \rho_0/(\mu^2 M^2), \phi_0 M_{pl}/M^2)$, where $\phi_0$ is the vacuum expectation value of the field and $\rho_0$ is the background cosmological density at $z=0$.

The $f(R)$ model was proposed by \cite{2007PhRvD..76f4004H} and can be defined with the following action
\be
S = \int \sqrt{-g} \left[ \frac{R+f(R)}{16\pi G} + \mathcal{L}_m \right] d^4 x, 
\ee
where $\mathcal{L}_m$ is the matter Lagrangian and the function $f$ is chosen as
\be
f(R) = - m^2\frac{c_1(R/m^2)^n}{c_2(R/m^2)^n+1}, 
\ee
where $m^2 \equiv H_0^2\Omega_m$ and $c_1$, $c_2$ and $n$ are dimensionless model parameters.  The number of parameters can be reduced to two by requiring the model to give $\Lambda$.  This requirement translates into $c_1/c_2 = 6\Omega_\Lambda/\Omega_m$.  As we did in the symmetron case, we redefine the parameters for numerical convenience and define 
\be
f_{R0} = -n\frac{c_1}{c_2^2}\left(\frac{\Omega_m}{3(\Omega_m+4\Omega_\Lambda)}\right)^{n+1}.
\ee
So we end up with two free parameters $f_{R0}$ and $n$.  The model can be written as a scalar tensor theory by making a conformal transformation of the metric \citep{brax}.  In this case, the scalar field fulfils an equation that has the same form as the chameleon field \citep[][]{{cham}}.  Thus, we will use the model as representative of the chameleon screening mechanism.  This mechanism is not associated with a specific symmetry as in the symmetron model, but works by increasing the mass of the scalar field.

Our analysis is based on a set of dark matter only cosmological simulations that were run with standard and modified gravity using the static modified gravity solver of the code Isis \citep{2014A&A...562A..78L}, which is a MG version of the GR state-of-the-art code RAMSES \citep{2002A&A...385..337T}.  The impact of non-static effects in these simulations was studied by \citet[][]{2013PhRvL.110p1101L, 2014PhRvD..89h4023L,2015JCAP...02..034B,2014PhRvD..89b3521N} and was found to be negligible for both models. The GR and MG simulations were run with the same initial seed, which means any differences between the runs arises from the difference in the gravity model.

In order to calculate covariance matrices and study the effects of our changes in the expected observational errors, we added 100 more realizations of the GR simulation to the original set of simulations.  Furthermore, we run a set of 50 symmetron simulations to confirm that the covariance matrices are independent of the gravitational theory.  The box size and number of particles used is 256 Mpc/$h$ and $512^3$ respectivelly.  The background cosmology is flat, given by $(\Omega_m, \Omega_{\Lambda}, H_0)=(0.267,0.733,71.9 ~\mathrm{km/sec/Mpc})$.  As we are interested in models that do not deviate much from GR, we only use the two simulations that display the weakest MG effects from the original set of simulations.  These are the simulations symm\_A and fofr6 in \citet{2014A&A...562A..78L}, for which the MG parameters are $(\lambda_0, a_{SSB}, \beta)=(1 ~\mathrm{Mpc}/h, 0.5, 1)$ and $(f_{R0},n)=(10^{-6},1)$.

\begin{versiona}
\begin{figure*} 
  \begin{center}
    \includegraphics[width=0.9\textwidth]{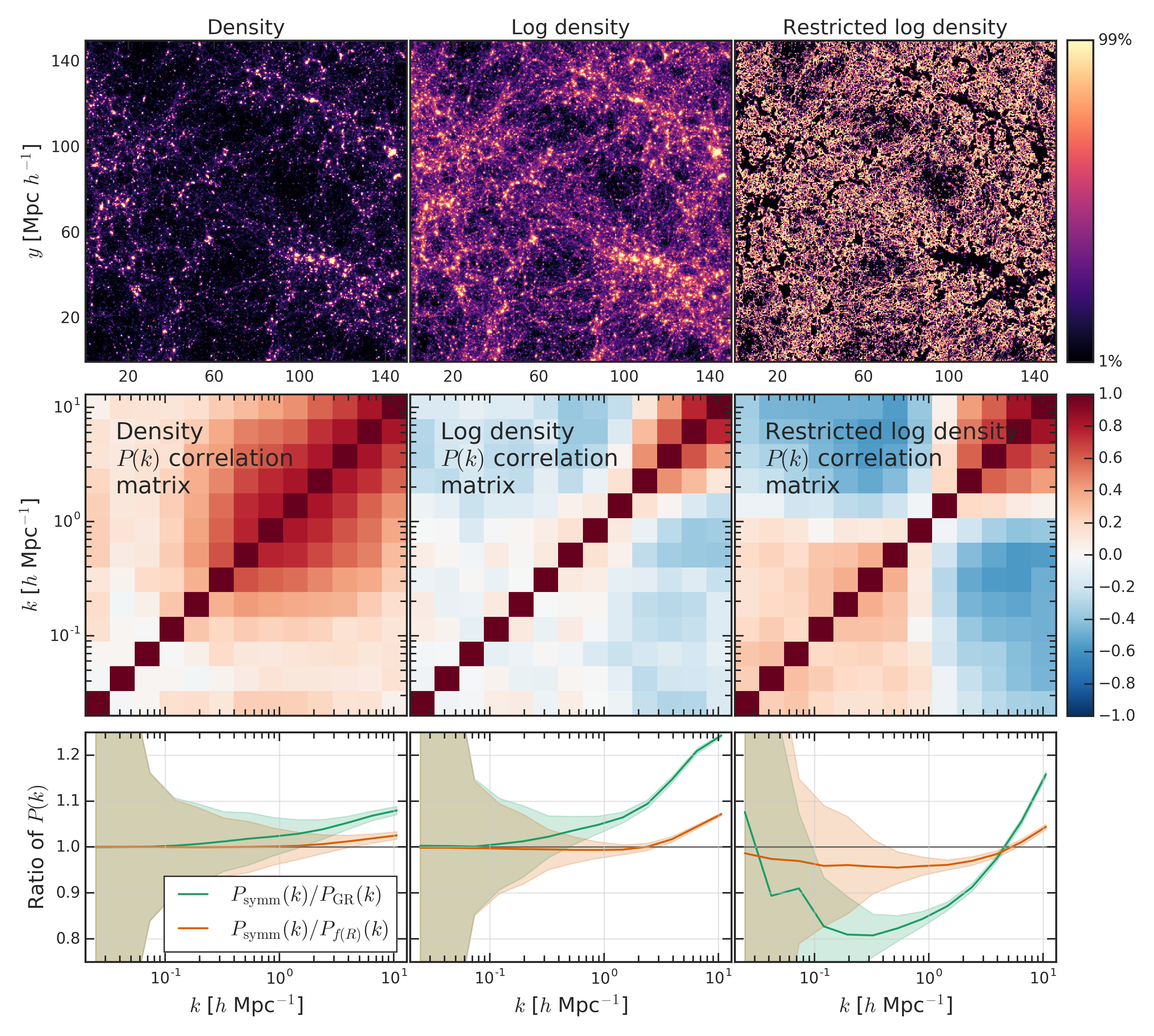}
    \caption{Top panels:  projected density field for each of the three transforms in GR.  Middle panels: associated correlation matrix of the power spectrum calculated using 300 GR maps.  Bottom panels:  ratio between symmetron and GR simulations (green) and between symmetron and $f(R)$ simulations (orange).  The shaded regions correspond to one standard deviation (see text for explanation).}
    \label{fig:lots_of_panels}
  \end{center}
\end{figure*}
\end{versiona}

\section{Analysis and results}

In order to prove the effectiveness of different weighting functions, we calculate the power spectrum of density perturbations for transformed density fields using the three transformations given by equations \ref{change_0} to \ref{change_2}.  We use 2D projections of 3D matter fields that are calculated by interpolating particle information into a uniform grid with 1024 nodes per dimension using a CIC kernel \citep{1988csup.book.....H}.  Three projections are extracted from each simulation, which correspond to the three Cartesian axes.  The power spectrum is computed via Fast Fourier Transform (FFT) of the 2D transformed density fields, and binning the power into 13 uniform logarithmic bins in frequency space.
The complete analysis pipeline can be found in \texttt{https://github.com/nualamccullagh/mg-log-xform/}.

Fig.~\ref{fig:lots_of_panels} shows our results.  Each column corresponds to one of the transformations of the density given by equations \ref{change_0} to \ref{change_2}.  The three free parameters of the restricted log density transform are fixed to $(A, B,C)=(3.01, -0.13,0.07)$.  These values were chosen by searching on a grid of parameters to find those that give both a large difference between the models \textit{and} a large signal-to-noise. The upper row shows an example of the transformed density fields from the GR box.  All three columns show the same region of the map.  The colour scheme is linear, and the figures show quantities between the 1st and 99th percentile of each distribution. The left-most column shows the usual density distribution, which is what is used in the standard definition of the power spectrum of density perturbations.  Here it is clear that the power spectrum is very sensitive to high-density peaks and thus contains limited information about modified gravity.  The log transform (middle column) reduces the importance of the halos, making the underlying cosmic web and internal structure of the voids more visible.  The right-most panel shows an example of the restricted log transform in which both the voids and halos are down-weighted and all the weight is given to an intermediate range of densities.

The second row in Fig.~\ref{fig:lots_of_panels} shows the normalized covariance matrices of the power spectra of the transformed density fields which are calculated using 300 maps extracted from the extended set of simulations described above (3 projections for each of the 100 GR realizations).  The power spectrum of the original density distribution (far left panel) shows large covariance on all scales, which is reduced by the log transform (middle panel).  This is no more than a confirmation of the results presented in \citet{2009ApJ...698L..90N}.  The covariance of the restricted log density transform (right panel), has positive off-diagonal terms at large and very small scales, and negative covariance between large and small scales.

The bottom row in Fig.~\ref{fig:lots_of_panels} shows the deviation between the symmetron and GR (green) and between the symmetron and $f(R)$ spectra (orange).  We use the GR covariances to compute 1-sigma error bars for these ratios (shaded regions), making the assumption that the covariance does not depend strongly on the gravity model. We check this assumption using the 50 symmetron realisations and find only negligible differences in the covariances of GR and symmetron. The fact that the error bars grow towards large scales is related to the limited number of modes near the fundamental mode of the box. We expect that the error on these scales would decrease with a larger volume. On small and intermediate scales, it is clear that both the log transform and the restricted log transform reduce the error bars compared to the usual density power spectrum. The curves in these panels correspond to the mean that was obtained using the three different projections taken from the original simulations.

Comparing the left and central panels shows the potential of the log transform to discriminate between GR and MG.  The log transform increases the difference between GR and symmetron models at small scales from values that were below 10\% for the usual power spectrum of the density field, to a difference that is above 20\% at $k\approx 10$ $h$/Mpc. Based on the error estimates shown, the deviation from GR appears to be significant (i.e. beyond two sigma) for $k\gtrsim 3$ $h$/Mpc for the usual power spectrum, and $k\gtrsim 1$ $h$/Mpc for the log density power spectrum. The log transform also amplifies the difference between symmetron and $f(R)$ from 2\% to 6\% at $k\approx 10$ $h$/Mpc; the deviation between the models becomes significant for $k\gtrsim 4$ $h$/Mpc for the log transform as opposed to $k\gtrsim 7$ $h$/Mpc for the usual power spectrum.  In both comparisons, the differences are restricted to very small scales, where predictions will certainly be affected by baryonic physics.

\begin{versiona}
\begin{table} 
  \centering
  \caption{Signal-to-noise of the overall power spectrum calculated from 300 GR maps and standard deviation of the ratio between different transformed power spectra at fixed angular frequencies.}
  \label{tab:results}
  \begin{tabular}{r | r r r }
    \thead{Mapping}                                & \thead{$f_0$} & \thead{$f_1$} & \thead{$f_2$} \strut  \\
    \hline 
    \thead{S/N (GR)}                                       & 85.80 & 434.60 & 485.67 \strut \\
   \hline 
    \thead{$\sigma(P_{symm}/P_{GR}; k=0.88 ~ h/\mathrm{Mpc})$}   & 0.083 & 0.039 & 0.033 \strut \\
    \thead{$\sigma(P_{symm}/P_{GR}; k=10.6 ~ h/\mathrm{Mpc})$}   & 0.018 & 0.008 & 0.013 \strut \\
   \hline 
    \thead{$\sigma(P_{symm}/P_{f_(R)}; k=0.88 ~ h/\mathrm{Mpc})$}        & 0.081 & 0.036  & 0.040  \strut \\
    \thead{$\sigma(P_{symm}/P_{f(R)}; k=10.6 ~ h/\mathrm{Mpc})$}        & 0.017  & 0.007  & 0.011 \strut 
  \end{tabular}
\end{table}
\end{versiona}

The differences become even larger when the log transform is replaced by the restricted log transform.  In this case, the overall difference in between the GR and symmetron power spectra has been amplified to 35\%.  Furthermore, the ratio starts to diverge from unity at larger scales: around $k\approx 0.2$ $h$/Mpc for distinguishing GR and symmetron.  The maximum deviation occurs around $k\approx0.3$ $h$/Mpc.  Moreover, the deviation between GR and MG is not monotonic as in the original density field or the log transform, but contains a feature which consists of a negative difference at scales larger than $k\sim 5~h/\mathrm{Mpc}$, which becomes positive for smaller scales.

The restricted log transform not only increases the difference between GR and symmetron, but also reveals differences between the two MG models which were hidden in the usual analysis.  In this case, the symmetron and $f(R)$ power spectra deviate from each other by more than two sigma with different signs at large and small scales.  The fact that the green and orange curves cross unity at almost the same point is a coincidence associated with the particular parameters used in this example.  Different parameters give different crossing points, and these do not always occur at the same scale for the two ratios shown.

To show that the proposed mappings do not increase the errors, we present in Table~\ref{tab:results} several quantities associated with the transforms.  The first line shows the signal-to-noise ratio of the overall power spectrum: $(S/N)^2 = \sum_{i,j}^{N_{\mathrm{bins}}} P(k_i) C_{ij}^{-1} P(k_j)$, which was calculated using the covariance from the 300 GR maps.  The remaining lines show the standard deviation associated with the ratios of the spectra between different models (which correspond to the shaded regions in Fig.~\ref{fig:lots_of_panels}) at specific frequencies.  In all the cases, our transforms increase the signal-to-noise and reduce the errors, making a potential detection more feasible.

\section{Conclusions}

The power spectrum of density perturbations (and 2-point statistics in general) is a standard tool used to constrain gravity and cosmological models.  While this is the simplest statistic associated with the matter distribution, and one of the main focuses of several of the next generation galaxy surveys, it is subject to a number of problems.  Firstly, taking into account Solar System observations when constraining MG models notably decreases the impact of MG on the power spectrum and thus our chances of detecting departures from GR.  Secondly, the effect of MG appears at scales where the effects of baryons and neutrinos are significant and thus, predictions are difficult and degenerate.  Finally, even in the case that we find departures from GR in the data, we still need to identify which alternative model is responsible for them.  Thus we need to find a way of maximizing the differences not only in the predictions between GR and MG, but also between different MG models.

Here we propose a solution to the first problem above, which could also help in alleviating the others and thus restore the suitability of the power spectrum as a probe for gravity.  This is crucial because several degeneracies affect the analysis, so the more independent observables are used, the better the chances of detecting new physics.  The method consists of calculating the power spectrum not of the density field, but of a transformed field, in which all the regions not affected by MG are down-weighted.  In the case of theories which include screening mechanisms (which are required to satisfy Solar System constraints), this can be done by reducing the weight of the high density regions.  We explore a more optimal transformation, which reduces the weight of both very high and very low density regions.

Our analysis is based on the symmetron and $f(R)$ MG simulations presented in \citet{2014A&A...562A..78L}.  As we are interested in realistic models, which do not depart much from GR, we choose the weakest fifth force models presented in that paper.  We found that a transformation of the density field exists (given by equation~\ref{change_2}) for which the differences between GR and MG can increase from less than 10\% to more than 35\%.  Furthermore, the difference between the two MG models increases from less than 3\% for the usual analysis to around 10\% for the same change.  Moreover, the domain of wave numbers where the models depart from each other shifts towards larger scales, into the regime where the baryons are sub-dominant.  Finally, the difference between different models is not monotonic as for the usual density field, but contains a feature, which will be difficult to reproduce using a linear bias and thus can help to disentangle between MG and baryonic effects. We note that this particular transform has been tuned for our simulations and the optimal parameters may depend on resolution and grid size.  This means that the determination of the best parameters has to be done taking into account the particulars of the target survey.

By analysing 100 GR and 50 symmetron realizations of the same boxes, we showed that the changes not only increase the differences between the models, but also increase the overall signal-to-noise of the power spectrum.  So the transformations presented here increase the detectability of MG models by both increasing the differences in the predictions and reducing the error bars.

Our study can be generalized in different ways.  Here we showed transformations that are targeted towards a possible detection of MG.  However, the same technique can be used to highlight different effects, such as baryonic physics, neutrinos, different dark matter models or baryon acoustic oscillations.  Different physical effects should have different functions associated to them.  By using different functions, it may be possible to break degeneracies for instance in between neutrinos and MG \citep{2014MNRAS.440...75B}.  This can be done by picking two different transformations that are sensitive only to one of these effects.  Finally, the simplicity of the method makes it ideal in comparison to similar ways of extracting non-Gaussian information from the data such as higher order statistics.  The method can also be generalized with respect to the data sets used.  Here we show the impact of transformations in the power spectrum of density perturbations.  However, a similar analysis can be done for instance with the velocity divergence power spectrum.

When it comes to parametric functions (such as the restricted log density presented in eq.\ref{change_2}), it can be useful not only to find a set of parameters that maximize the efficiency of the transformation, but to study the behaviour of the predictions when parameters of the transformation are changed.  This can give extra information about the models, which can not be obtained from the usual 2-point statistics.

Strict solar system constraints make the detection of MG signatures in cosmological data as difficult as looking for a needle in a haystack.  We present a novel technique which sweeps away the straws and keeps only the needles, increasing our chances of a near future detection.

\section*{Acknowledgements}

We thank Carlton Baugh, Shaun Cole, Mark Neyrinck and Peder Norberg for helpful discussions.  Special thanks to Lydia Heck for support with computer issues.  
This work used the DiRAC Data Centric system at Durham University.  This equipment was funded by BIS National E-infrastructure capital grant ST/K00042X/1, STFC capital grants ST/H008519/1 and ST/K00087X/1, STFC DiRAC Operations grant ST/K003267/1 and Durham University. DiRAC is part of the National E-Infrastructure.  We acknowledge support from STFC consolidated grant ST/L00075X/1 \& ST/P000541/1.  NM acknowledge support from European Research Council Grant (DEGAS-259586).

\bibliographystyle{mnras}
\bibliography{references}


\bsp    
\label{lastpage}

\end{document}